\documentclass[10pt, conference, compsocconf]{IEEEtran}
\ifCLASSINFOpdf
  \usepackage[pdftex]{graphicx}
\else
  \usepackage[dvips]{graphicx}
\fi
\usepackage{tabularx}

\usepackage{comment}

\ifCLASSOPTIONcompsoc
\usepackage[caption=false,font=normalsize,labelfont=sf,textfont=sf]{subfig}
\else
  \usepackage[caption=false,font=footnotesize]{subfig}
\fi
\usepackage[T1]{fontenc} 
\usepackage{amsmath}

\usepackage[cmintegrals]{newtxmath}
\usepackage{bm} 
\usepackage{cite}
\usepackage{xcolor}
\def\BibTeX{{\rm B\kern-.05em{\sc i\kern-.025em b}\kern-.08em
    T\kern-.1667em\lower.7ex\hbox{E}\kern-.125emX}}
\usepackage{placeins}
\usepackage{diagbox}
\usepackage[inline]{enumitem}
\usepackage{fancyhdr} 
\usepackage{ragged2e} 
\fancypagestyle{IEEEcopyright}{
    \fancyfoot[L]{\justifying \footnotesize 
    © 2025 IEEE. Personal use of this material is permitted. Permission from IEEE must be obtained for all other uses, in any current or future media, including reprinting/republishing this material for advertising or promotional purposes, creating new collective works, for resale or redistribution to servers or lists, or reuse of any copyrighted component of this work in other works. This paper is accepted to appear in the TrustSense workshop of PerCom'25.
    }
}

\begin{document}
%
\title{On the Impact of Sybil-based Attacks on Mobile Crowdsensing for Transportation}



\author{\IEEEauthorblockN{Alexander Söderhäll}
\IEEEauthorblockA{Networked Systems Security Group \\
KTH Royal Institute of Technology \\
Stockholm, Sweden \\
Email: asoderh@kth.se}
\and
\IEEEauthorblockN{Zahra Alimadadi}
\IEEEauthorblockA{Networked Systems Security Group \\
KTH Royal Institute of Technology \\
Stockholm, Sweden \\
Email: alimadad@kth.se}
\and
\IEEEauthorblockN{Panos Papadimitratos}
\IEEEauthorblockA{Networked Systems Security Group \\
KTH Royal Institute of Technology \\
Stockholm, Sweden \\
Email: papadim@kth.se}}


%


\maketitle
\thispagestyle{IEEEcopyright} 

\begin{abstract}

Mobile Crowd-Sensing (MCS) enables users with personal mobile devices (PMDs) to gain information on their surroundings. Users collect and contribute data on different phenomena using their PMD sensors, and the MCS system processes this data to extract valuable information for end users. Navigation MCS-based applications (N-MCS) are prevalent and important for transportation: users share their location and speed while driving and, in return, find efficient routes to their destinations. However, N-MCS are currently vulnerable to malicious contributors, often termed Sybils: submitting falsified data, seemingly from many devices that are not truly present on target roads, falsely reporting congestion when there is none, thus changing the road status the N-MCS infers. The attack effect is that the N-MCS returns suboptimal routes to users, causing late arrival and, overall, deteriorating road traffic flow. We investigate exactly the impact of Sybil-based attacks on N-MCS: we design an N-MCS system that offers efficient routing on top of the vehicular simulator SUMO, using the InTAS road network as our scenario. We design experiments attacking an individual N-MCS user as well as a larger population of users, selecting the adversary targets based on graph-theoretical arguments. Our experiments show that the resources required for a successful attack depend on the location of the attack (i.e., the surrounding road network and traffic) and the extent of Sybil contributed data for the targeted road(s). We demonstrate that Sybil attacks can alter the route of N-MCS users, increasing average travel time by 20\% with Sybils 3\% of the N-MCS user population.
\end{abstract}



%
\IEEEpeerreviewmaketitle

\section{Introduction}
Networking technologies, such as 5G and Wi-Fi, allow Personal Mobile Devices (PMDs) to receive and transmit large amounts of data. This has boosted Mobile Crowd-Sensing (MCS) systems. An MCS system collects location-based data leveraging PMD sensors, notably including Global Navigation Satellite System (GNSS) receivers. Navigation MCS-based applications (N-MCS), such as Google Maps\footnote{https://www.google.com/maps} and Waze\footnote{https://www.waze.com/live-map}, recruit users to share their location and speed data when on the road. This data allows the N-MCS to estimate road traffic state and provide users with efficient routes by computing the Expected Time of Arrival (ETA)~\cite{GoogleETA}, among other things. 

For an MCS system to provide a service, it must recruit users whose PMDs collect data. The most straightforward recruitment method is to make the MCS app easy to use, often only requiring an email address to register and collect data\footnote{For example, GMap: https://support.google.com/accounts/answer/27441}. This makes MCS-based apps vulnerable to malicious users, typically termed Sybil-based adversaries, performing data falsification attacks. A Sybil-based attack on an N-MCS system (for brevity, a Sybil attack from now on) is an attack where an adversary uses multiple user accounts (i.e., acting as an insider) and coordinates them to submit falsified location data and, by extension, speed. 
Commercial N-MCS systems (Google Maps and Waze) were successfully Sybil attacked and misled in traffic state estimation. These attacks were successful against real-world N-MCS~\cite{SybGoogle,WangWWNZZ:J:2018,wazeblackhat}, affecting targeted areas without the adversary physically deploying devices in those areas. Thus, a tangible threat, especially as the adversary can scale the data falsification attack, increasing adversarial data volume and affecting large parts of the road network. Manipulating the N-MCS to create the illusion of congestion implies that the average speed in affected road segments is computed lower than the actual one. Thus, any ETA calculated based on data poisoned by the Sybils is higher than the actual one, leading benign users to select alternative routes and experience longer travel times. More benign users following suboptimal routes can also lead to their actual congestion.

However, no research has quantified the impact of Sybil attacks on benign N-MCS users. Clearly, there are ethical constraints: one cannot attack real N-MCS systems to affect benign users present. Still, the impact on benign users is of great interest, which is why this paper, same as in ~\cite{SHIEELD,VCS}, uses a realistic micro-mobility traffic simulator (Simulation of Urban MObility (SUMO)~\cite{SUMO2018}): we construct a simplified (explained in Sec.~\ref{design}) yet robust N-MCS that provides efficient routing to its users. The adversary model uses graph theory, explained in Sec.~\ref{sim}, to determine high-value roads. Sybil attacks are mounted against an individual N-MCS user and a larger population of users, with different number of Sybils, varying speeds, attack durations, and target roads with differing features, conducted at different times of the day (Sec.~\ref{res}). 
Our investigation confirms that N-MCS users indeed avoid Sybil attack targets, with Sybils being 3\% of the N-MCS user population, and this can increase travel time by up to 20\% on average. An adversary is most effective/successful when Sybils target multiple vital routes simultaneously and report the lowest speed possible. On the other hand, the more benign users traversing the target roads during the attack, the lower the attack impact. Our contributions are:
\begin{enumerate}
    \item Simplified, efficient N-MCS on top of SUMO.
    \item Adversarial targeting strategy based on Betweenness Centrality (BC) measurements.
    \item Quantified impact of Sybil attacks on benign N-MCS (notably, user travel time and time loss) and analysis of determining factors (road network topology and attacker strength). 
\end{enumerate}

The rest of this paper is organized as follows: Section~\ref{rel} outlines background and related work. Sec.~\ref{problem} states the problem and Sec.~\ref{sys} details the system and adversary models. The proposed method is explained in Sec.~\ref{des}, and experimental results in Sec.~\ref{res}, before we conclude.
\section{Background \& Related Work} \label{rel}
A general MCS system has:
\begin{enumerate*}[label=(\roman*)] 
\item 
users collecting and contributing data on specific tasks, 
\item task initiators defining sensing tasks (the sensing objective, region, and duration, the reporting frequency, and the number of samples per report), and 
\item 
infrastructure (i.e., servers) that processes data~\cite{TrustPeople}.
\end{enumerate*}
 N-MCS systems require continuous reporting and live data processing; thus, the infrastructure uses a windowing mechanism to process live stream data as in~\cite{Sboing}. The windowing mechanism has two parameters:
\begin{enumerate*}[label=(\roman*)] 
\item window size ($w.size$), determining the time interval for which window holds data for processing, i.e., [-$w.size$, now] and
\item window slide ($w.slide$), determining how frequently the window should update to the current time. 
\end{enumerate*} The window can either be time-based or size-based. As traffic data is time-sensitive, N-MCS systems use time-based windowing. 
Using a window for data processing facilitates responsiveness as new data is collected. Historical data allows an N-MCS to use seasonal and historical trends of road traffic when computing the ETA~\cite{Sboing}. Different N-MCS use different approaches: GMap~\cite{GoogleETA} uses a graph neural network (GNN)-based model, while Sboing~\cite{Sboing} uses a linear regression-based model. As there is a lack of historical data for our experiments, we only consider window-based steaming processing. Moreover, GNN-based traffic prediction models are also vulnerable to evasion attacks, where the adversary inserts malicious data to mislead the model at inference time~\cite{DiffusionGNN}, the same analogy as the Sybil attack. Thus, being the first paper exploring the impact of the Sybil attack on benign N-MCS users, we use Dijkstra~\cite{dijkdijk}, which is the foundation of most modern routing algorithms, for ETA computation based on window-based steaming processing.

The recent Sybil attacks on commercial N-MCS~\cite{SybGoogle,WangWWNZZ:J:2018,wazeblackhat} exploit their openness and lack of verification of the contributing PMD positions. They implement the attacks using emulators or scripts that conduct man-in-the-middle for data falsification, which are scalable. Different methods have been proposed to detect Sybils in N-MCS. Eryonucu et al.~\cite{MCSSybil} propose a collaborative scheme for position verification to counter Sybils. Wang et al.~\cite{P2PSybil} model the problem as community detection to detect Sybil devices based on colocation of devices. Yu ~\cite{sybiltraj} utilizes a generative-based model to detect Sybils, assuming Sybil trajectories are synthetic rather than actual, physical PMD movement. To the best of our knowledge, commercial N-MCS systems have not yet implemented any prevention or detection approach. For example, Sboing~\cite{Sboing} only uses streaming distance-based outlier detection, which can be defeated by an attacker that controls enough data points to form a valid cluster in a given window. In this work, we investigate the impact of Sybils on benign users and are in control of them. Thus, the detection of Sybils is orthogonal to this work.

\section{Problem Statement}\label{problem}
Despite the vulnerability of the N-MCS systems to Sybil attacks, no research has quantified their impact on benign users, notably having Sybil attackers submitting faulty data with a lower speed in a particular location, creating an illusion of road traffic congestion. Thus, in an N-MCS without any mechanism for filtering out these malicious data, artificially and adversely reduced speed will lead the N-MCS to compute incorrectly higher travel times. The N-MCS computed shortest path will likely avoid the artificially congested road(s). But in reality, the user may end up taking a longer route in terms of travel time than the route it would have taken in the absence of the attack. The question is \emph{how does a Sybil attack on N-MCS impact transportation}, and \emph{what are the contributing factors for a successful attack}? 

\section{System and Adversary Model}\label{sys}
We consider a single-task N-MCS with a server, without loss of generality, also playing the role of task initiator. The task requires users to submit speed and location data every second. The server uses a windowing mechanism, with size ($w.size$) and slide ($w.slide$) as its configuration parameters, to process the data. It uses the submitted speed and location data to create a live view of the traffic and provide the users with the shortest path (time-wise). Without loss of generality, we assume that the map matching is done on the application side when collecting data\footnote{SUMO simulation provides information on the road the cars drive on, which makes map matching straightforward.}. Moreover, the N-MCS is assumed to be designed with security in mind, preventing external adversaries from eavesdropping and tampering with the data of other users in transit using typical end-to-end encryption such as TLS and user authentication. The N-MCS system server has a correct road network layout and information with all road speed limits, lengths, and connectivity.

All N-MCS users, benign or Sybil, are assumed to adhere to the task-specific data transmission frequency.  Submission of higher-rate data can trivially be ignored by resampling each user stream of data. The adversary (i.e., Sybil attacker) aims to pollute the N-MCS system by sending falsified data, reflecting a falsified view of road traffic congestion. The ultimate goal of the adversary is to alter the Expected
Time of Arrival (ETA) of benign users and degrade vehicle routing efficiency; which could consequently undermine the N-MCS system.
To do so, the adversary registers multiple authenticated accounts and coordinates them to conduct attacks, exploiting N-MCS openness and, in some cases, reusing authentication tokens. An adversary can coordinate Sybils using script-based or emulation-based approaches~\cite{SybGoogle}, effectively enabling the adversary to spawn the (fictitious) Sybils at a location of choice and submit data as if they travel along any road at any target speed. 
\section{Simulation based N-MCS}\label{des}
Our approach is to experimentally quantify the impact of Sybil attacks on benign users using a micro-mobility vehicular simulator. First, we select a simulator that allows live modification of vehicle movement. We then emulate the N-MCS functionality, and based on the computed shortest path it would provide to the user, we reroute N-MCS user vehicles. In this setup, we simulate attacks of different strengths to evaluate the effect of the attacks quantitatively. Sec.~\ref{sumo} introduces the chosen simulator and the base urban mobility we used for our experiments. Sec.~\ref{design} explains how we emulate N-MCS, and Sec.~\ref{sim} describes the simulation of the attacks. Finally, Sec.~\ref{metrics} discusses the evaluation parameters.

\subsection{Simulation of an Urban Environment}\label{sumo}

We use Simulation of Urban MObility (SUMO)~\cite{SUMO2018}, an open-source micro-mobility traffic simulator that allows vehicle speed and location data to be retrieved and vehicle routes modified during a live simulation using a built-in library named TraCI\footnote{Traffic Control Interface: https://sumo.dlr.de/docs/TraCI.html}, in multiple pre-made SUMO scenarios\footnote{https://sumo.dlr.de/docs/Data/Scenarios.html}. To enable realistic traffic and road network situations, we selected the InTAS road network, simulating 187,500 vehicles in Ingolstadt, Germany, over 24 hours~\cite{INTAS}. InTAS has approximate optimal routing through Dynamic Traffic Assignment (DTA), approximating optimal vehicle routing by running simulations and iteratively selecting more efficient routes for a subset of the vehicle population until no more efficient routes can be detected. Route efficiency is based on three metrics:
\begin{enumerate*}[label=(\roman*)] 
\item mean vehicle speed, 
\item time loss\footnote{Time loss is the difference between the theoretical time to travel the route given optimal conditions and the actual time to travel the route.}, and
\item travel time (i.e., routing duration in SUMO context).
\end{enumerate*} 

\subsection{Emulation of N-MCS}\label{design}

We emulate an N-MCS system on top of SUMO/TraCI using Python, implementing: 
 \begin{itemize}
\item{\textbf{N-MCS Users}}: These are selected vehicles based on the N-MCS user penetration rate (defined as the percentage of the total), as in the real world, not all vehicles contribute to an N-MCS. These are randomly chosen among vehicles (with predefined trip start and destination) in the InTAS simulation. 

\item{\textbf{N-MCS Application}}: TraCI facilitates an approximate implementation of an N-MCS app, retrieving vehicle data and updating vehicle routes based on the ETA. All users send data as per the task reporting frequency (i.e., every 1 $s$): 
\begin{equation}
    R_{i,t} =  \{id_{i}, l_{i}, s_{i}, t\}
\end{equation}

$R_{i,t}$ is the report submitted by user $i$ at timestep $t$, $id_{i}$ is a user identifier (assigned by SUMO), $l_{i}$ the lane (i.e., road) the user is currently at, and $s_{i}$ is the speed, in $m/s$, the user is traveling at.

\item{\textbf{N-MCS Server}}: With a time-based streaming window, similar to Sboing~\cite{Sboing}, the travel time of each road is estimated by taking the mean over the data in the window and dividing it by the road length. To reduce computation overhead, for each road, the speed is computed as the mean of data submitted at the same timestep, not waiting for the entire window to fill before computing the average over all data in the window. Then, the road speed is estimated as the average speed of these averages in a window. The N-MCS stores the speed estimates and allows users to find the fastest route from their current position to their destination (using Dijkstra~\cite{dijkdijk} on the graph representing the road network and speed estimates). Dijkstra offers a clear computation for the N-MCS provided, which, based on extensive simulation results (Sec.~\ref{res}, Fig.~\ref{slide}) without attacks, performs better than InTAS DTA in SUMO.
\end{itemize}

\subsection{Simulation of the Attacks}\label{sim}

A Sybil attack is a coordinated attack with multiple authentic N-MCS user accounts. The adversary falsifies the location of the accounts/"users" it controls and, by extension, the speed; we simulate the attacks beforehand in a separate empty InTAS simulation. Therefore, similar to the real-world implementation of such an attack, Sybil attackers are not physically present at the attack location. During the live simulation of the attack, we feed their data reports into the N-MCS data streaming window, reflecting congestion on the targeted road(s). Given one or multiple targeted roads, we insert the Sybils driving with an attack \textit{speed} at the start of the target road(s). Sybil \textit{speed} is set using the TraCI function $\emph{traci.vehicle.setMaxSpeed()}$. New Sybils are inserted into the road once enough space exists, and they utilize all available lanes of a target road, driving along until they exit the road.

The Sybil adversary can use different strategies to select the target road(s). Through publicly available data (e.g., OpenStreetMap\footnote{ https://www.openstreetmap.org}), the adversary can model the road network as a weighted directed graph by knowing the speed limits, the length of all roads, and their connectivity. 
Therefore, given the graph representation of the road network, the adversary can compute the Betweenness Centrality (BC) to deduce possible vital road targets. BC is defined as per Equation~\ref{BC11}~\cite{BC1}.

\begin{equation}
    c_B(v) = \sum_{s,t\in V}\frac{\sigma(s,t|v)}{\sigma(s,t)}
    \label{BC11}
\end{equation}

$V$ is the set of all the nodes in the road network, $c_B(v)$ is the BC of a node $v \in V$. $\sigma(s,t)$  is the number of shortest routes for a given source ($s$) and target ($t$) location, and $\sigma(s,t|v)$ denotes the number of shortest routes passing through $v$. We visually inspect the proposed targets of \textit{BC} for a meaningful selection. A road is meaningful to attack if it has high neighboring connectivity or it is a bottleneck. For example, highways tend to get high scores based on \textit{BC}. However, the highway may be the only option for vehicles to enter the city, given their starting position. Therefore, we avoided them based on visual inspection. More broadly, successful attack strategies depend on road connectivity and traffic flow, and there can be all kinds of optimality criteria and trade-offs. Investigation of attacker strategies can be part of future work.

\subsection{Evaluation Metrics}\label{metrics}
We first evaluate the routing efficiency of our N-MCS in comparison to the base InTAS DTA simulation using \emph{time loss}, and \emph{travel time}, considering different user penetration rates and select N-MCS window, $w$, parameters accordingly. Then, having determined the efficiency of our N-MCS system, we quantify the impact of the Sybil attack on \textit{an individual N-MCS user} and \textit{an N-MCS population subset}. The experiments are designed as follows:

\begin{enumerate}[label=(\alph*)]

\item \textbf{Impact on an Individual N-MCS User}:
We determine the contributing factors to the success of a Sybil attack on an individual N-MCS user. We add a vehicle to the original SUMO scenario, which is the only user following the N-MCS routing (i.e., the route(s) N-MCS returns), while the rest of the N-MCS users only contribute data. We consider target road(s) where the Sybils are (not) the only contributors, altering the \textit{number} of Sybils and their \textit{speed} and observing if the victim avoids the targeted roads by the adversary. The Sybil attack is performed before the victim starts, ensuring that all Sybil data are in the N-MCS window. 

\item \textbf{Impact on an N-MCS Population Subset}:

We quantify the attack impact on an N-MCS population subset by computing the \emph{time loss} and \emph{travel time}.
We conduct the same attack during different \textit{times of the day} and vary the attack \textit{duration} to account for different traffic flows.
We run our N-MCS without any Sybils present as our baseline and then use the same seed as the baseline to simulate our attacks\footnote{It should be noted that all the simulations have the same vehicles as N-MCS, which were randomly selected in the beginning.}. We then identify the 
set of affected N-MCS users as those who used the target road(s) during the attack period in the baseline scenario. Comparing the traces for these affected users in baseline and under attack, we define two sets:  
\emph{Did Enter}; if they still enter the attack road target(s) and \emph{Did Not Enter}; if they avoid the road target(s). For the set \textit{Did Not Enter}, the evaluation does not concern their entire route, only the difference in their route. The difference in their routes is determined by finding the first common starting road shared between the simulations with and without a Sybil attack present, tracking backward from the Sybil attack target road(s). The same procedure is repeated to find the first common ending road.
\end{enumerate}

\section{Experimental Results}\label{res}

\begin{figure}[!t]
    \centering
    \includegraphics[width=0.999\linewidth]{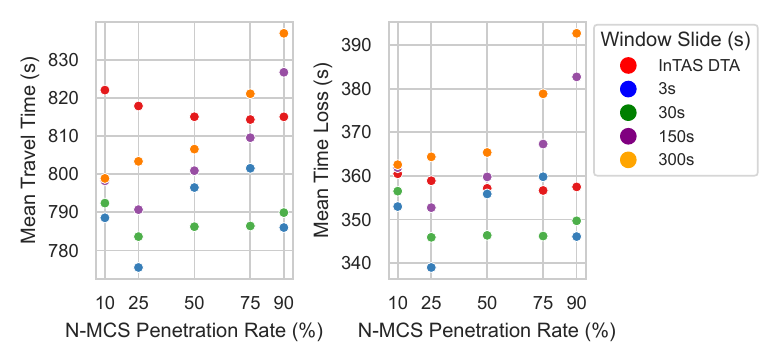}
    \caption{Comparison of the impact of the N-MCS window slide on the routing efficiency of our N-MCS, with default InTAS using DTA.}
    \label{slide}
\end{figure}

To evaluate the routing efficiency of our N-MCS, as per the configuration of  Sboing~\cite{Sboing}, we set $w.size$ to 300$s$ and found the best $w.slide$ among $\{3, 30, 150, 300\}$ in comparison to the InTAS DTA, considering different N-MCS user penetration rate in $\{10, 25, 50, 75, 90\}$. Fig.~\ref{slide} shows the efficiency of our N-MCS routing compared to the InTAS DTA simulation. A $w.slide$ 30$s$ provided the best routing evaluation parameters (lower mean for both time loss and travel time) across all MCS populations, showing better results than InTAS DTA, all other settings of $w.slide$ and a lower variance between penetration rates than 3$s$. Thus, we set the $w.slide$ as 30$s$ for the rest of the experiments. The N-MCS penetration rate is set to 50\% of the vehicles as N-MCS users, as this allows a realistic model of an actual percentage of N-MCS users in traffic\footnote{https://www.statista.com/statistics/432169/online-route-planning-and-map-usage-eg-google-maps-germany}.

\subsection{Impact on an Individual N-MCS User}\label{victim}

\begin{table}[!tb]
  \begin{center}
    \caption{Sybil attack strength when targeting an N-MCS user.}
    \label{SybilStrength}
    \begin{tabular}{|c|c|} 
    \hline
        Parameter & Values \\
        \hline
        Number of Sybils & \{2, 4, 6, 8, 10, 12, 14, 16\} \\
        Faulty data: Speed (m/s) & \{0.5, 1, 1.5, 2, 3, 4, 5, 6\} \\
        \hline
    \end{tabular}
  \end{center}
\end{table}

\begin{figure*}[t]
\centering

\begin{minipage}[c][8.5cm][t]{\textwidth}
    \begin{minipage}[c][4cm][t]{0.49\textwidth}
    \vspace{1.2cm}
        \subfloat[Target I: 1 Sybil, Speed: 1 $m/s$]{
                    \includegraphics[width=\textwidth]{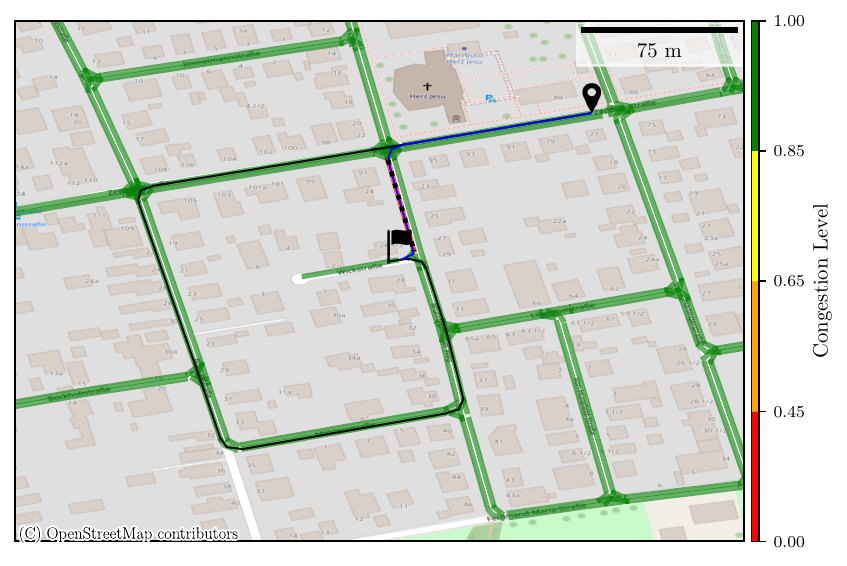}
        \label{attackI}}
    \end{minipage}%
    \hfill
    \begin{minipage}[c][4cm][t]{0.49\textwidth}
        \subfloat[Target II: 1 Sybil, Speed: 6 $m/s$]{
                    \includegraphics[width=\textwidth]{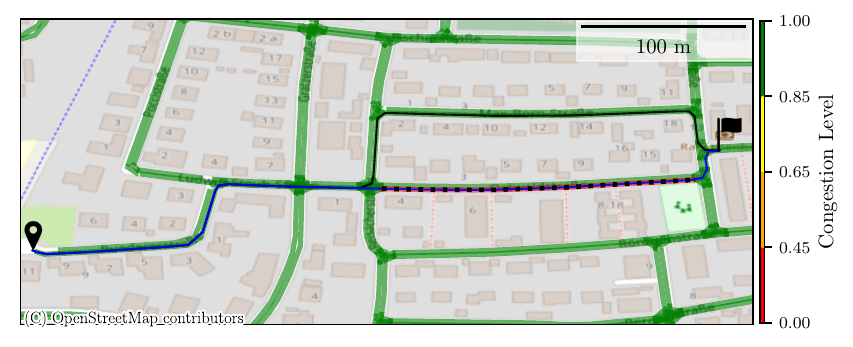}
        \label{attackII}}
        \vspace{-0.3cm}
        \subfloat[Target III: 1 Sybil, Speed: 2 $m/s$]{
                    \includegraphics[width=\textwidth]{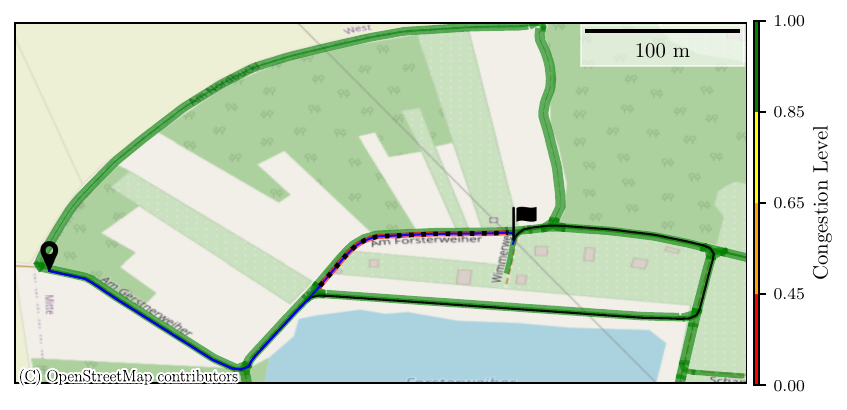}
        \label{attackIII}}
                
    \end{minipage}
\end{minipage}

    \begin{minipage}[c][7.5cm][t]{\textwidth}
        \centering
            \subfloat[Target IV: 4 Sybils, Speed: 6 $m/s$]{
                \includegraphics[width=0.28\textwidth]{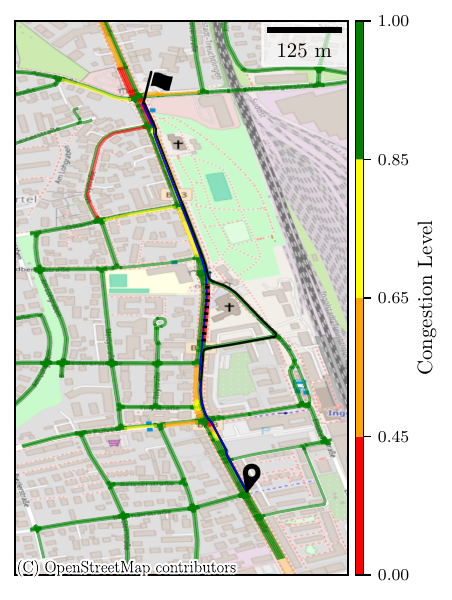}
            \label{attackIV}}
            \hfill
            \subfloat[Target V: 6 Sybils, Speed: 0.5 $m/s$]{
                \includegraphics[width=0.28\textwidth]{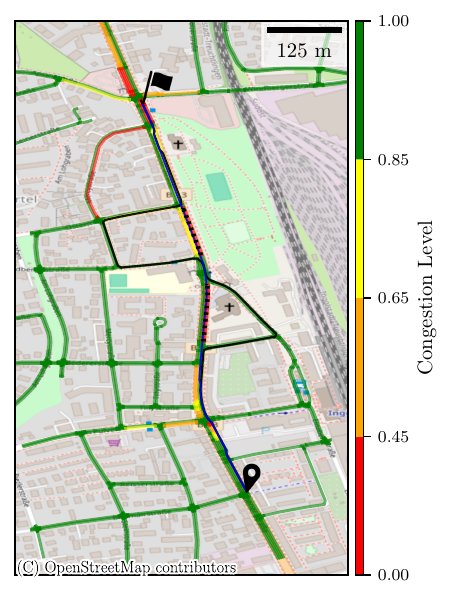}
            \label{attackV}}
            \hfill
            \subfloat[Target VI: 12 Sybils, Speed: 0.5 $m/s$]{
                \includegraphics[width=0.28\textwidth]{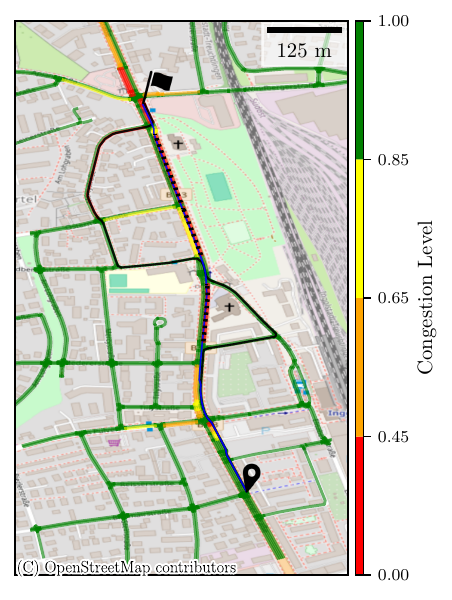}
            \label{attackVI}}
            
    \end{minipage}

\caption{Visualization of the Sybil attacks on an individual N-MCS user starting from \protect\includegraphics[width=0.5em]{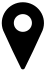} with a destination of \protect\includegraphics[width=0.5em]{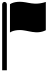}, with the traffic state as when the reroute is triggered. The original route of the victim is marked in blue, while the deviation from the original route during the Sybil attack (black dotted lines) is in black. The number of Sybils in each attack is the minimum required per target road, driving at the maximum required speed for a successful re-routing. Fig.~\ref{attackI}-~\ref{attackIII} are the Sybil attacks where no Benign N-MCS users were on the target roads, while Fig.~\ref{attackIV}-~\ref{attackVI} are those with Benign N-MCS users presence.} 
    \label{fig:sybilattackvictim}
\end{figure*}

To find the contributing factor for a successful Sybil attack, triggering a re-route of an individual N-MCS user (i.e., victim), we consider target roads with and without benign users traversing them during the Sybil attacks, with different strengths as in Table~\ref{SybilStrength}. Three target roads (I, II, and III) were selected as no benign N-MCS users traverse these roads along the victim trip (shown with the dotted line in Figs.~\ref{attackI}-~\ref{attackIII}). These roads were randomly selected by visual inspection, considering the road was empty during the victim trip. The results show that Sybil \textit{speed} is the only important parameter in defining Sybil strength when adversarial data ("users") are alone on the target road (see Figs.~\ref{attackI}-\ref{attackIII}). 

Another three roads with benign N-MCS users were chosen as adversary targets through analysis using BC. They form a cluster of roads along the same route, thus referred to as \emph{clustered} road targets. Three different combinations of these roads (targets IV, V, and VI) were selected, as shown in green in Fig.~\ref{fig:targets}. We set our victim trip with a start and destination, resulting in the shortest path containing this clustered road in a benign scenario (see Figs.~\ref{attackIV}-\ref{attackVI}). 
Our results show that when benign N-MCS users are present on the target roads, the number of Sybils also plays an important role. In other words, the presence of benign users increases the adversary resources required for a successful attack (Figs.~\ref{attackIV}-\ref{attackVI}).

Given enough resources, the Sybil attacks altered the victim route in all scenarios, as shown in Fig.~\ref{fig:sybilattackvictim}. Effectively, to trigger a re-route, the ETA of the original route must exceed that of the second-best route. Thus, the ETA of the second-best route determines the required Sybil resources to deploy. The number of Sybils only mattered when the target road(s) had benign N-MCS users traversing them. This is expected as benign user contributions counteract adversarial data thus increasing resources to alter the ETA. We also observe that the resource requirements for re-routing a vehicle from multiple roads are not linear. As the Sybil attacks occurred during the morning rush hour, traffic increased in the surrounding area (as the traffic state shows in Figs.~\ref{attackIV}-\ref{attackVI}). 
This increased the ETA of all roads, thus increasing the adversary resource requirements.

\subsection{Impact on an N-MCS Population Subset}

\begin{figure}[!tb]
    \centering\includegraphics[width=\linewidth]{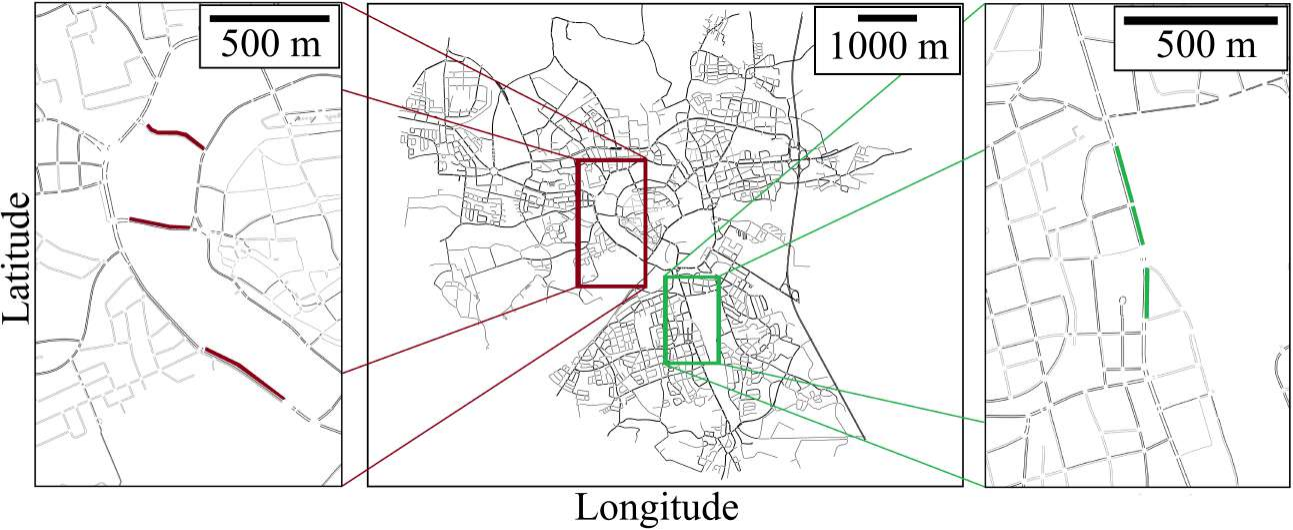}
    \caption{The location of clustered targets (in green) and multi-routed targets (in red). For \textit{clustered targets}, with a direction towards the north: Target IV comprises the bottom, green-colored road, Target V the bottom two, and Target VI all three. For \textit{multi-routed targets}, with a direction towards the west: Target VII comprises the bottom, red-colored road, Target VIII the bottom two, and Target IX all three.}
    \label{fig:targets}
\end{figure}

To quantify the Sybil attack impact on a subset of the N-MCS population, we use the clustered targets (road segments IV, V, and VI, in green in Fig.~\ref{fig:targets}) and add three new targets using the adversary target analysis based on BC. These new targets, also shown in Fig.~\ref{fig:targets} but in red, referred to as \emph{multi-route targets}, comprise road(s) on the three most efficient routes from the southern part to the west and northwest parts of the InTAS map. We selected three different combinations of these roads for simulating the attacks: Target VII is part of the most efficient road, Target IX of the two most efficient ones, and Target VIII of all three roads. 
Given our attack simulation method, Table~\ref{rr1} outlines the \textit{number} of Sybils for each target for three different attack \textit{durations}. The Sybil attacks were conducted at three \textit{different times of day}: \emph{morning rush hour}, \emph{afternoon rush hour}, and \emph{calm evening}. We set the attack \textit{speed} to 0.5 $m/s$ for all experiments, based on the result for Sybil attacks on an Individual N-MCS user.

\begin{table*}[t]
\vspace{1.05cm}    
    \centering
    \caption{Total (maximum active) Sybil attack strength per duration when targeting an N-MCS population subset.}
    \begin{tabular}{|c||c|c|c||c|c|c|}
    \hline
    \diagbox[width=4cm]{Duration (min)}{Target} & IV & V & VI & VII & VIII & IX \\
        \hline
        20 & 67 (37) & 144 (69) & 208 (112) & 14 (14) & 36 (36)  & 67 (59) \\
        40 & 166 (38) & 348 (69) & 517 (112) & 117 (95) & 191 (126) & 274 (149) \\
        60  & 263 (38) & 551 (69) & 825 (112) & 222 (96) & 349 (128) & 485 (151)  \\
        \hline
    \end{tabular}
    \label{rr1}
\end{table*}

\begin{figure}[t]
\centering
\subfloat[Sample Ratio (\%)]{\includegraphics[width=0.5\textwidth]{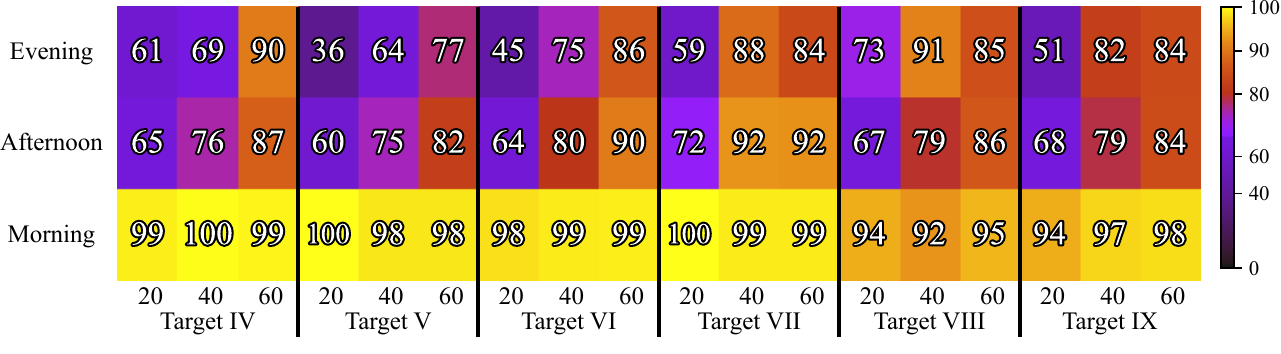}
\label{fig:sample_ratio}}
\vfill
\subfloat[Travel Time (\%)]{\includegraphics[width=0.5\textwidth]{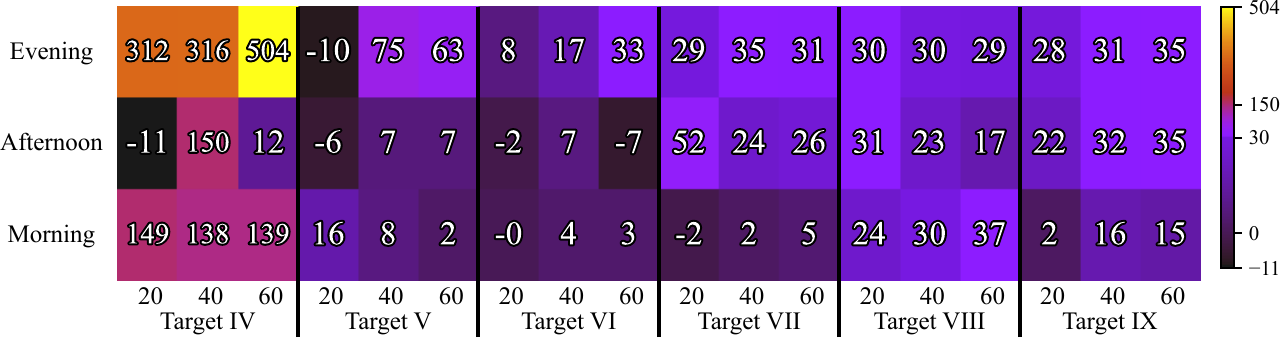}
\label{fig:duration}}
\vfill
\subfloat[Time Loss (\%)]{\includegraphics[width=0.5\textwidth]{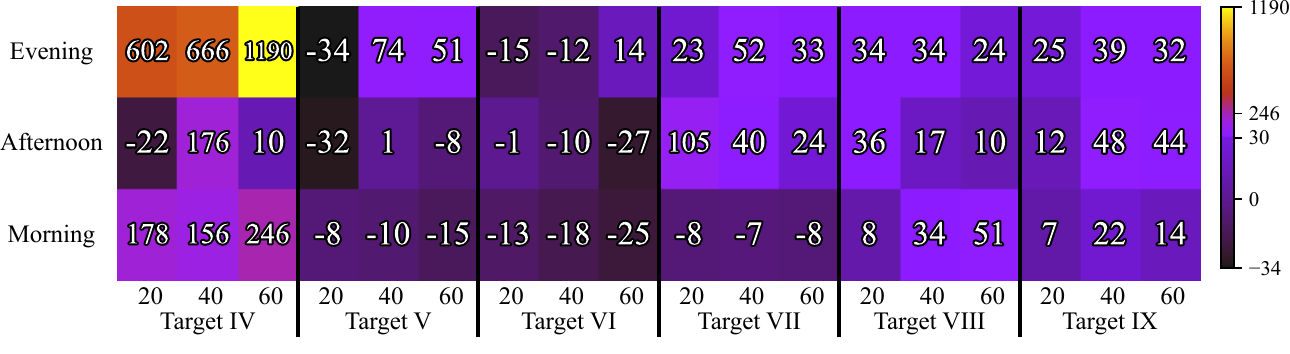}
\label{fig:tloss}}

\caption{\ref{fig:sample_ratio} shows the ratio of the \emph{Did Not Enter} set compared to all affected N-MCS users for an attack.~\ref{fig:duration} and~\ref{fig:tloss} show the median percentage change of the metrics for the \textit{Did Not Enter} set across road targets, considering Sybil attack durations (in min) over different times of the day.}
    \label{fig:quntification}
\end{figure}

\begin{figure}[tb]
  \centering
\includegraphics[width=0.5\textwidth]{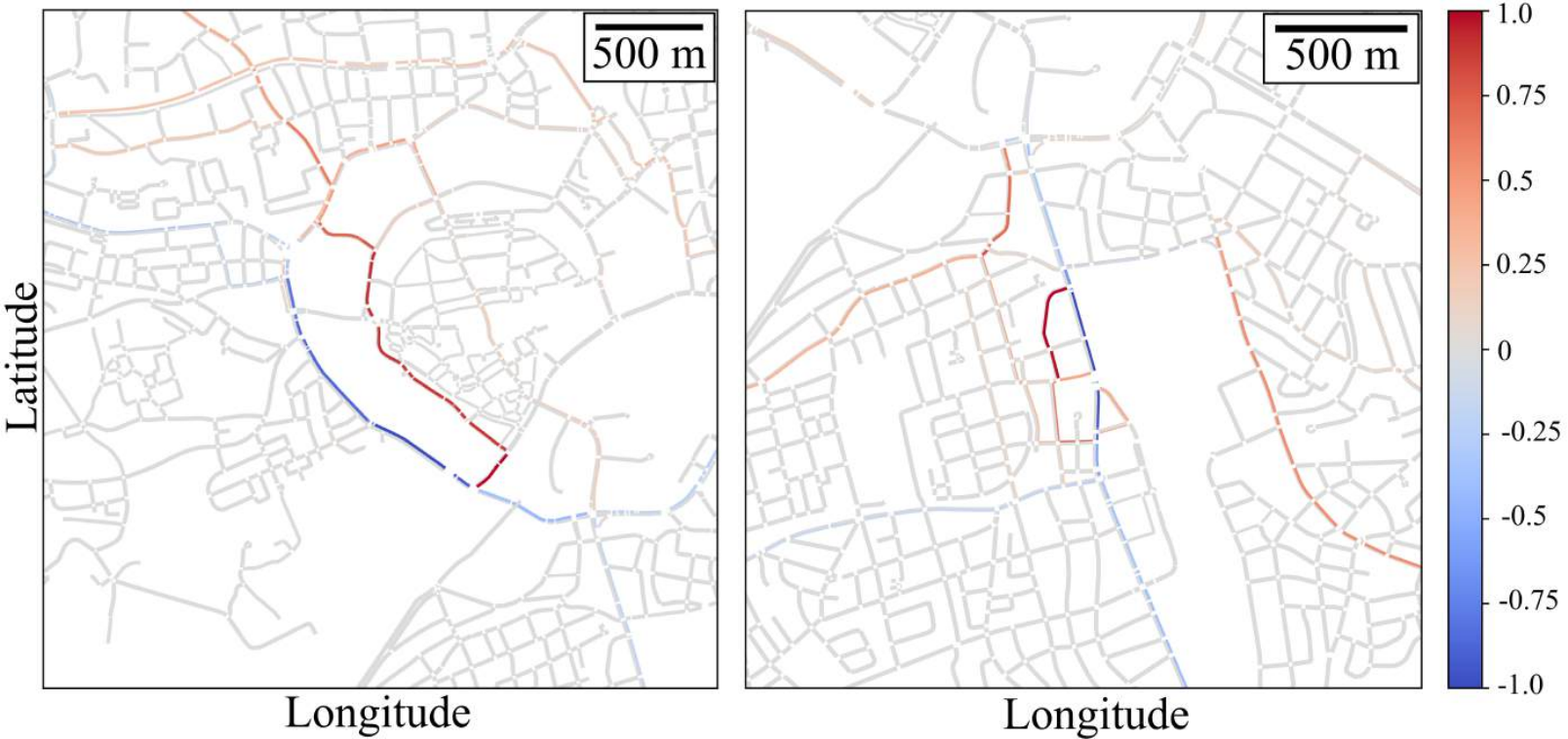}
  \caption{Change in the traffic flow of the affected N-MCS users by a Sybil attack during calm evening traffic. 
  Blue/Red indicates a decrease/increase in the number of N-MCS users entering the roads due to the attack. Left/Right shows the impact of Sybil attack on Target VIII/VI lasting for 20/60 min. The darkest red road had 37/110 more benign users traversing it, and the darkest blue road had 63/302 fewer benign users traversing it during a Sybil attack.}
    \label{rrs}
\end{figure}

For the set \textit{Did Enter}, vehicles that continued entering the target road(s) during a Sybil attack, across all 54 unique Sybil attacks, we saw no noticeable metric trends. 
However, for the set \textit{Did Not Enter}, which diverged from the target road(s) during a Sybil attack, we saw noticeable effects, due to distinct differences in vehicle/user routing. Fig.~\ref{fig:quntification} visualizes the impact of all 54 unique Sybil attacks on set \textit{Did Not Enter}. Fig.~\ref{fig:sample_ratio} shows the ratio of the set \textit{Did Not Enter} to all the affected N-MCS users defined as:
\begin{equation}
    \text{Sample Ratio} = \frac{|\textit{Did Not Enter}|}{|\textit{Did Not Enter}| + |\textit{Did Enter}|}
\end{equation}
During the \textit{morning rush hour}, most of the affected N-MCS users were rerouted in all experiments. During the \textit{afternoon rush hour} and \textit{calm evening} attacks, the longer the duration of an attack, the more successful the attack becomes, rerouting a higher majority. In general, the shortest attack during the \textit{calm evening} was the least impactful one in terms of the percentage of rerouted vehicles in all locations. We visualize the impact of these reroutes on the flow of affected N-MCS users for Target VI and VIII with an attack duration of 60 $min$ and 20 $min$, respectively, in Fig.~\ref{rrs}, traversing suboptimal routes, and possibly creating actual congestion. Regarding the impact on travel time, as shown in Fig.~\ref{fig:duration}, the \emph{clustered} targets experience higher variance than the \emph{multi-route} targets, highlighting the importance of the neighboring road network. Furthermore, attacks during \emph{morning rush hour} had, in general, lower impact in comparison to \emph{afternoon rush hour} and \emph{calm evening}, reflecting that the state of traffic also plays a part, as morning traffic flowing into the city takes the reverse direction in the evening.

The impact is also captured by the time loss shown in Fig.~\ref{fig:tloss}: for multi-routed targets: most of the Sybil attacks negatively impact the \textit{Did Not Enter} set; attacks on Target VII during the \textit{morning rush hour} being an exception. For Target VII during the \textit{morning rush hour} there is a decrease in time loss, meaning vehicles mostly drove with speed closer to the ideal speed. The same can be seen for Target V and VI, while for Target IV, we see a higher variance, which is the same as its travel time metric. 

\renewcommand{\thesubfigure}{\alph{subfigure}}

Across all attacks on the N-MCS subset population, the Sybils were, on average, 76\% of the N-MCS users that traversed the target road(s) during the attack but only approximately 3\% of the N-MCS population during the attacks. Given that a sizeable number of Sybils need to be "deployed" when benign N-MCS users are present, if an attack succeeds in re-routing a significant fraction of benign users away from the target, the adversary can lower their number of Sybils on the target road and still falsify congestion on the road.  
Calculating the mean difference in travel time across all 54 Sybil attacks on the N-MCS subset population, taking into account the sample size of each set \emph{Did Not Enter}, the Sybil attacks, although only 3\% of the N-MCS population, increased travel time by 20\% on average. 

The attack impact on the \textit{clustered} and \textit{multi-routed} targets differs significantly, likely due to the road network near the clustered target providing accessible re-routing opportunities. 
To increase the attack impact, the adversary should "deploy" its resources (Sybils) to attack multiple routes and attempt to contest vital road segments, such as those leading from one area to another. This is observable in our results on the \textit{multi-routed} targets, with attacks more predictably increasing travel time (compared to \emph{clustered} targets). 

From a different perspective, the increased time loss showcases that the attack-induced routes did not manage to handle larger amounts of traffic, with the affected N-MCS users experiencing significant time loss. Live data could not predict the congestion, and it is questionable whether historical data would have sufficed. To increase resilience and robustness, N-MCS systems should use the characteristics of the road network to predict the outcome of traffic flow, decreasing the chance that Sybil attacks can cause significant road congestion due to diverting traffic using ill-fit roads.

\section{Conclusion}\label{disc}

We implemented a simple but robust N-MCS system on top of SUMO, which allowed us to investigate the impact of Sybil attacks on benign users for the first time. For the adversary target road selection, we used the \textit{BC} algorithm, identifying the vital roads in urban areas. Our results showed that a multitude of parameters determine the success, impact, and resource requirements of Sybil attacks: strategic and possibly majority "placement" of Sybils on targeted roads, the road network itself, and the current traffic conditions. A successful Sybil attack can increase the travel time of N-MCS users on average by 20\%, even though the attacker may control Sybils that amount to no more than 3\% of the N-MCS user population. This highlights the importance of devising defense mechanisms to thwart Sybil attacks against N-MCS.



\bibliographystyle{IEEEtran}

\bibliography{IEEEabrv,bibliography}

\enlargethispage{1in}

\end{document}